\documentclass[twocolumn,final]{svjour2}

\usepackage{latexsym,amsmath}
\usepackage{amsbsy}
\usepackage[pdftex]{graphicx}
\usepackage{amssymb}
\usepackage{epstopdf}
\usepackage[usenames]{color}
\usepackage{float}
\usepackage{graphicx}
\usepackage[normalem]{ulem}

\begin{document}

\title{Soft communities in similarity space}

\author{Guillermo Garc\'{i}a-P\'erez \and M. \'Angeles Serrano \and Mari\'an Bogu\~n\'a}
\institute{Guillermo Garc\'{i}a-P\'erez \at
Departament de F{\'i}sica de la Mat\`eria Condensada, Universitat de Barcelona, Mart\'{i} i Franqu\`es 1, 08028 Barcelona, Spain \\ 
Universitat de Barcelona Institute of Complex Systems (UBICS), Universitat de Barcelona, Barcelona, Spain \\
\email{guille.garcia@ub.edu} \and
M. \'Angeles Serrano \at
Departament de F{\'i}sica de la Mat\`eria Condensada, Universitat de Barcelona, Mart\'{i} i Franqu\`es 1, 08028 Barcelona, Spain \\
Universitat de Barcelona Institute of Complex Systems (UBICS), Universitat de Barcelona, Barcelona, Spain \\
ICREA, Pg. Llu\'is Companys 23, E-08010 Barcelona, Spain \\\email{marian.serrano@ub.edu} \and
Mari\'an Bogu\~n\'a \at
Departament de F{\'i}sica de la Mat\`eria Condensada, Universitat de Barcelona, Mart\'{i} i Franqu\`es 1, 08028 Barcelona, Spain \\
Universitat de Barcelona Institute of Complex Systems (UBICS), Universitat de Barcelona, Barcelona, Spain \\
\email{marian.boguna@ub.edu}}

\titlerunning{Soft communities in similarity space}
\authorrunning{G. Garc\'{i}a-P\'{e}rez, M. Bogu\~{n}\'{a} and M. \'{A}. Serrano}

\maketitle

\begin{abstract}
The $\mathbb{S}^1$ model has been a central geometric model in the development of the field of network geometry. It has been mainly studied in its homogeneous regime, in which angular coordinates are independently and uniformly scattered on the circle. We now investigate if the model can generate networks with targeted topological features and soft communities, that is, heterogeneous angular distributions. Under these circumstances, hidden degrees must depend on angular coordinates and we propose a method to estimate them. We conclude that the model can be topologically invariant with respect to the soft-community structure. Our results might have important implications, both in expanding the scope of the model beyond the independent hidden variables limit and in the embedding of real-world networks.
\end{abstract}

\section{Introduction}
Complex networks have been widely studied in the last twenty years in many different contexts, from biological to social and technological~\cite{newmanbook,Dorogovtsev:2003ey}. There seems to be some universal features common to the topology of many networks. For instance, in most cases, they are scale-free, meaning that their degrees are power-law distributed, a phenomenon that was explained in early times of network theory by the preferential attachment mechanism: as the network grows, new nodes connect to highly connected ---or popular--- nodes with higher probability~\cite{Barabasi:1999ay}.

However, preferential attachment alone cannot explain the high level of clustering coefficient ---the fraction of existing triangles--- observed in real systems. To explain clustering, the concept of similarity was introduced~\cite{Serrano:2008ga}. The basic idea is that nodes connect not only because they are popular, but also because they are similar in some sense. Thus, if node $A$ connects to nodes $B$ and $C$ because they are similar to $A$, $B$ and $C$ should also be similar and therefore have a high probability of being connected. This \textit{transitivity of similarity} suggests encoding similarities between nodes as distances in metric spaces, since the triangle inequality is one of their defining properties: if the distance $d_{BC}$ in the underlying metric space measures the dissimilarity between $B$ and $C$, it must be bounded by $d_{BC} \leq d_{AB} + d_{AC}$, therefore inducing the observed transitive connections.

The $\mathbb{S}^1$ model was proposed based on these ideas~\cite{Serrano:2008ga}. In this model, $N$ nodes are randomly scattered into a circle of radius $R=N/2 \pi$. Every node $i$ is also assigned a hidden degree $\kappa_{i}$ from any distribution (for instance, a power-law $P(\kappa) \sim \kappa^{-\gamma}$), and every pair of nodes $i$ and $j$ is connected with probability 
\begin{equation}\label{eq:connection_probability}
p_{ij} = \frac{1}{1 + \left(\frac{d_{ij}}{\mu \kappa_{i} \kappa_{j}}\right)^{\beta}},
\end{equation}
where $d_{ij}$ is the distance along the circle; $\mu$ and $\beta$ are two global parameters controlling the average degree and the clustering coefficient, respectively. Notice that this connection probability takes the form of a gravity law, as it increases with the product of hidden degrees (popularities) and decreases with the distance between them (dissimilarity). Despite its apparent simplicity, this model generates networks that resemble very much real networks; they are scale-free, small-world and have high levels of clustering. In fact, the degrees $k_i$ are proportional to the hidden degrees $\kappa_i$, so the model is versatile enough to generate networks with different degree distributions\footnote{The original model defined in~\cite{Serrano:2008ga} is in fact more general, allowing for any connection probability $p_{ij}$ as long as it depends on the argument $d_{ij}/(\kappa_i \kappa_j)^{1/D}$, where now the space is the $D$-dimensional sphere and $d_{ij}$ the geodesic distance on the sphere. The particular functional form in Eq.~\eqref{eq:connection_probability} allows us to interpret the network as a set of non-interacting fermions (the links) embedded in the hyperbolic plane, with the hyperbolic length of a link playing the role of its energy and $\beta$ playing the role of the inverse of the bath temperature~\cite{Krioukov:2009aa}.}.

The possibilities of the $\mathbb{S}^1$ model go beyond generating realistic networks. The similarity space coordinates of the nodes of a real network can be inferred by finding the coordinates that maximise the likelihood for the real network to be generated by the model~\cite{Boguna2010,Serrano:2012we,Garcia-Perez:2016}. This embedding process yields a map of the network strikingly meaningful. For instance, it allows to navigate the network efficiently by mapping the coordinates to hyperbolic space. Moreover, being able to access to the similarity space coordinates of nodes opens the path to a completely new way of analysing complex networks. For example, in Refs.~\cite{Boguna2010,Serrano:2012we,Garcia-Perez:2016} it was shown that the angular coordinates of nodes in real-world networks are not uniformly distributed. Instead, they are distributed in a heterogeneous manner, with angular regions more densely populated than others. These dense regions reveal the community structure of the network~\cite{Newman:2004aa,Radicchi:2004av,Arenas:2008aa}. Indeed, by partitioning the network using the largest gaps between consecutive nodes along the circle as community boundaries, the partitions obtained have a modularity comparable to that of other community detection methods currently available in the literature. Furthermore, this geometric method seems to have higher resolution~\cite{Garcia-Perez:2016}.

In Ref.~\cite{Zuev:2015aa}, the authors introduced the Geometric Preferential Attachment (GPA) model, a generalised version of the growing geometric model Popularity vs. Similarity Optimization~\cite{Papadopoulos:2012uq} in which soft communities, as they named these denser angular regions, emerge from the growth dynamics of the network without altering topological properties like the degree distribution or the clustering spectrum of the resulting network. In this paper, we address the question of whether the $\mathbb{S}^1$ model can generate networks with given target topological features and soft communities, that is, \textit{heterogeneous} angular distributions. To that end, any angular distribution could be considered in principle. For instance, one could impose some non-uniform distribution function a priori. However, the angular distribution from Ref.~\cite{Zuev:2015aa} is not an imposition, but it rather emerges from a preferential attachment process in similarity space that seems to be a plausible explanation for the nature of communities in real systems. We focus on that particular angular distribution for our study.

\section{Results}

In the bare $\mathbb{S}^1$ model, hidden degrees and similarity coordinates are typically assumed to be uncorrelated, so every node's hidden variables are withdrawn independently from some joint distribution $\rho(\kappa, \theta)$ that factorises~\cite{Serrano:2008ga,KrPa10}. Nevertheless, the GPA is a growing model, so the angular coordinates and hidden degrees of different nodes are correlated. We are therefore forced to drop such simplifying assumptions.

In the GPA growth process, the degree of a node is determined by its age ---the older the node is, the higher its degree. Moreover, when a new node $t$ is added to the system, the probability for it to be placed at polar coordinate $\theta_t$ depends on the number of nodes $s < t$ at angular distance $\Delta \theta_{st} < 2 / (s^{\frac{1}{\gamma-1}} t^{\frac{\gamma-2}{\gamma-1}})$, where $\gamma$ is the exponent of the power-law degree distribution. This implies a very particular dependence between similarity coordinates and degrees: the angular coordinate of a node must depend on the angular coordinates of all nodes with higher degree. Hence, we must include the implicit ordering in the sequence of nodes induced by the degree sequence in our heterogeneous version of the $\mathbb{S}^1$ model. That could be done by first assigning a hidden degree $\kappa_i$ from a power-law distribution $P(\kappa) \sim \kappa^{-\gamma}$ to every node, ordering the nodes according to their hidden degrees and reproducing the angular preferential attachment from the GPA model with that particular ordering. At the end of the process, we would obtain a set of $N$ nodes with hidden degrees power-law distributed with exponent $\gamma$ and the same angular distribution as the GPA model for that value of $\gamma$. However, if we then connected every pair of nodes with the probabilities given by Eq.~\eqref{eq:connection_probability}, degrees and hidden degrees would not be proportional; the reason for such deviation from the usual behaviour of the model is that a homogeneous angular distribution is required for the proportionality between hidden and observed degrees~\cite{KrPa10}, which is not fulfilled here by construction; hidden degrees must depend on the spatial distribution of nodes as well. In the following subsection, we address this issue. We explore the heterogeneous regime of the $\mathbb{S}^1$ model and show that it is capable of generating networks with power-law degree distributions, high clustering and soft communities.

\subsection{Geometric Preferential Attachment in the $\mathbb{S}^1$ model}
From the previous discussion, we see that hidden degrees and angles are considerably entangled in the modelling of geometric networks with soft communities. In this context, the $\mathbb{S}^1$ model requires the following steps:

\begin{figure*}[t!]
\centering
\includegraphics[width=\linewidth]{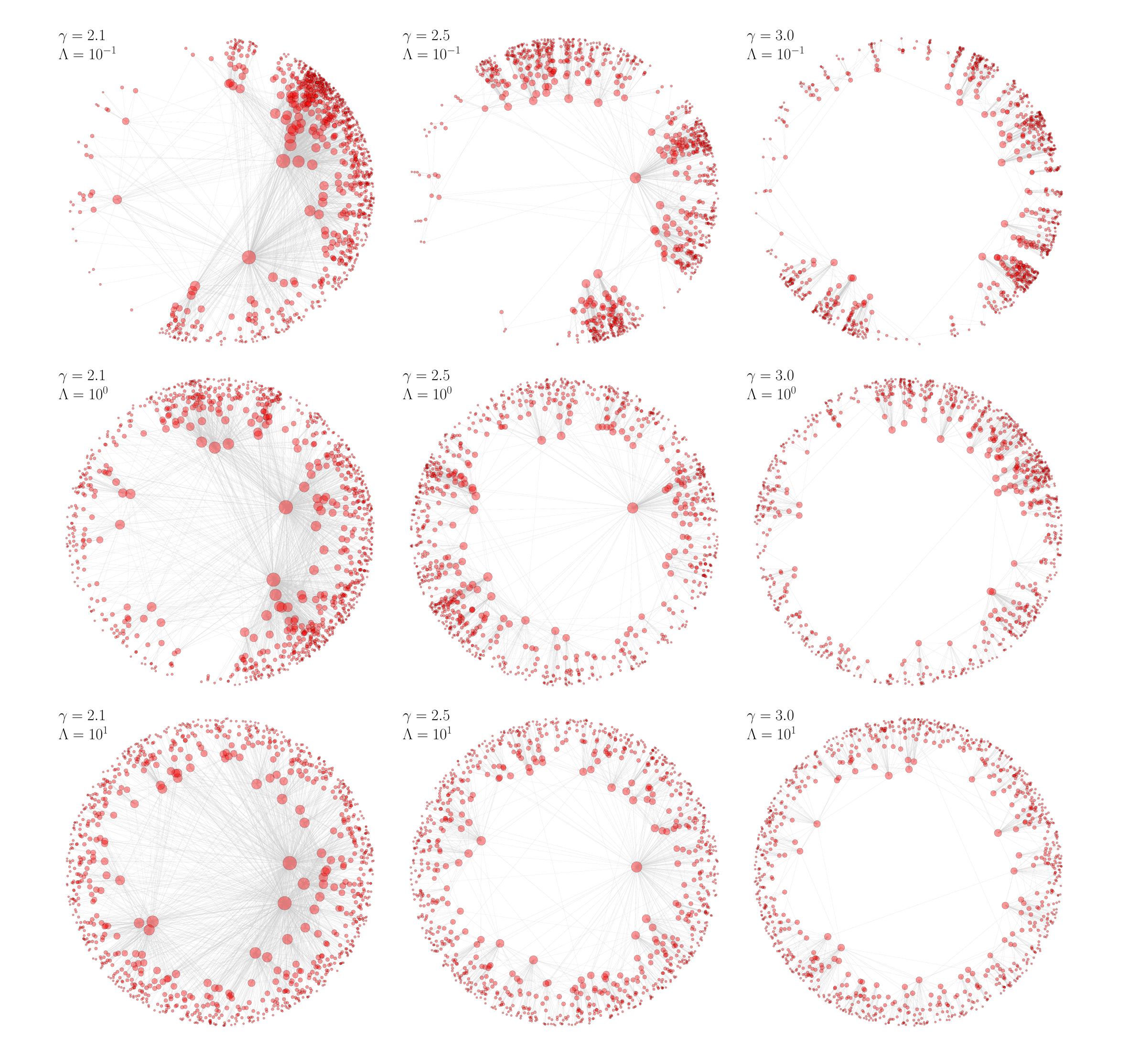}
\caption{\label{fig:angles} Geometric layout of the networks generated by the $\mathbb{S}^1$ model with Geometric Preferential Attachment. In all cases, $N=1000$ and $\beta = 2.5$. Every column corresponds to a value of $\gamma$ and every row to a value of $\Lambda$. As in Ref.~\cite{Zuev:2015aa}, soft communities emerge for low values of the initial attractiveness $\Lambda$. In order to clarify the figure, every node's target degree is represented as a radial coordinate $r_{i} = R - 2 \ln k_{i}^{\mathrm{tar}} / k_{N}^{\mathrm{tar}}$, where $k_{N}^{\mathrm{tar}}$ is the smallest target degree and $R = 2 \ln \left( N/(\pi \mu (k_{N}^{\mathrm{tar}})^2) \right)$. When using the hidden degrees instead of the target degrees, this mapping constitutes the isomorphism between the $\mathbb{S}^1$ model and the $\mathbb{H}^2$ model in hyperbolic space~\cite{KrPa10,Boguna2010,Serrano:2012we,Garcia-Perez:2016}.}
\end{figure*}

\begin{figure*}[t!]
\centering
\includegraphics[width=\linewidth]{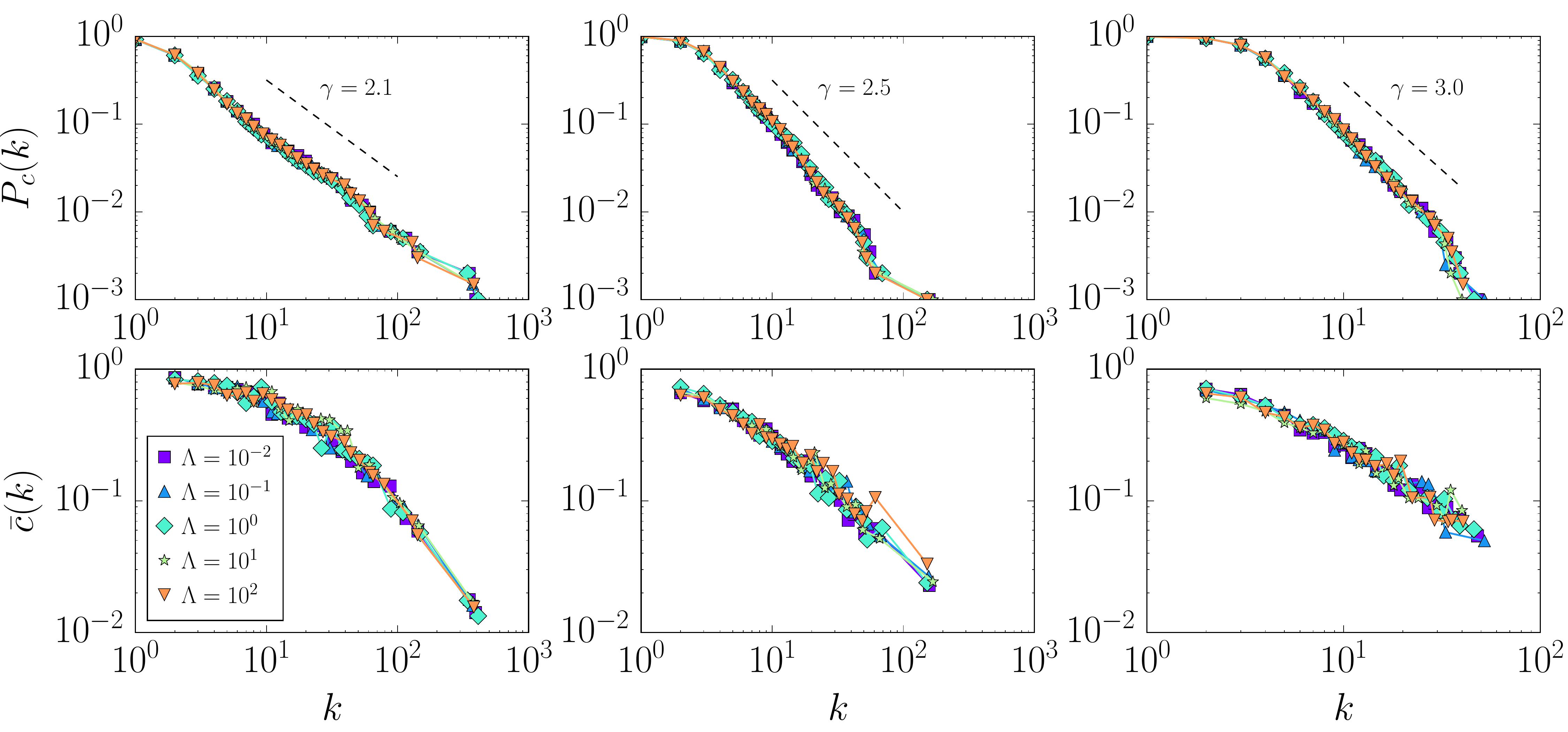}
\caption{\label{fig:topologies}Topological properties of the networks generated by the $\mathbb{S}^1$ model with Geometric Preferential Attachment with $N=1000$ and $\beta = 2.5$. Every color corresponds to a different value of the initial attractiveness $\Lambda$. The top row shows the complementary cumulative degree distribution $P_c(k) = \int_{k}^{\infty} P(k) \mathrm{d}k$, which behaves as $P_c(k) \sim k^{1-\gamma}$ for $P(k) \sim k^{-\gamma}$. Black dashed lines indicate such scaling. In the bottom row, the mean local clustering spectra $\bar{c}(k)$ are drawn. To avoid fluctuations in the target degrees, all networks with the same exponent $\gamma$ have been generated with the same target degree sequence $\lbrace k_{i}^{\mathrm{tar}} \rbrace$. Despite their angular distributions being completely different (Fig.~\ref{fig:angles}), their topologies are extremely similar.}
\end{figure*}

\begin{itemize}
\item[1.] \textbf{Assigning angular coordinates:} Angular coordinates are assigned according to the Geometric Preferential Attachment, which requires an ordering. Therefore, assign a label $i = 1,\ldots,N$ to every node. Then, for every value of $i$ from $1$ to $N$:

\begin{itemize}
\item[a.] Sample $i$ candidate angular positions $\phi_{j}, j=1,\ldots,i$ for node $i$ from $U(0, 2\pi)$.
\item[b.] For every candidate position, define the attractiveness $A(\phi_j)$ of candidate $j$ as the number of nodes $s$ with an already defined angular position, that is, with $s < i$, at angular distance $\Delta \theta_{is} < 2 /(s^{\frac{1}{\gamma-1}} i^{\frac{\gamma-2}{\gamma-1}})$, where $\gamma$ is the exponent of the target power-law degree distribution.
\item[c.] Assign to node $i$ the angular coordinate of candidate $j$, i.~e. set $\theta_i = \phi_j$, with probability
\begin{equation}\label{eq:assignment_probability}
\Pi(\phi_j) = \frac{A(\phi_j) + \Lambda}{\sum_{n=1}^{i} \left( A(\phi_n) + \Lambda \right)}.
\end{equation}
The initial attractiveness $\Lambda \geq 0$ is a parameter that sets the strength of the geometric preferential attachment. For very high values of $\Lambda$, all candidate angles become equally likely, so the resulting angular distribution is homogeneous (see Fig.~\ref{fig:angles}).
\end{itemize}
This process generates a distribution of nodes in the circle analogous to the angular distribution of the GPA model. However, notice that, in the GPA model, connections are established \textit{at the same time} as positions are decided, whereas in the former steps, no connections have yet been made.

\item[2.] \textbf{Assigning hidden degrees:} Once every node has a defined angular position, we need to determine its hidden degree such that the resulting observed degrees, that is, after the connections have been actually established, are power-law distributed with exponent $\gamma$. As mentioned earlier in this paper, we must take into account that the spatial distribution is heterogeneous (especially for low values of $\Lambda$). We propose the following method:

\begin{itemize}
\item[a.] Generate a set of $N$ target degrees $k^{\mathrm{tar}}$ from a power-law distribution with exponent $\gamma$. Order the target degrees such that $k_{1}^{\mathrm{tar}} > k_{2}^{\mathrm{tar}} > \cdots > k_{N}^{\mathrm{tar}}$.
\item[b.] Assign to every node $i$ a hidden degree $\kappa_i$, initially set to $\kappa_{i} = k_{i}^{\mathrm{tar}}$.
\item[c.] Repeat $N$ times:

\begin{itemize}
\item[i.] Choose some node $i$ randomly.
\item[ii.] Compute the expected degree $\bar{k}_{i}$ of node $i$ as
\begin{equation}\label{eq:expected_degree}
\bar{k}_{i} = \sum \limits_{j \neq i}  \frac{1}{1 + \left(\frac{d_{ij}}{\mu \kappa_{i} \kappa_{j}}\right)^{\beta}}.
\end{equation}
\item[iii.] Correct the value of $\kappa_{i}$ so that the expected degree $\bar{k}_{i}$ matches the target degree $k_{i}^{\mathrm{tar}}$. We propose to reset $| \kappa_{i} + \left( k_{i}^{\mathrm{tar}} - \bar{k}_{i} \right) \delta | \rightarrow \kappa_{i}$, where $\delta$ is a random variable withdrawn from the uniform distribution $U(0, 0.1)$. Other numerical methods could be used with the same end.
\end{itemize}
\item[d.] Compute all relative deviations
\begin{equation}\label{eq:relative_errors}
\epsilon_i = \frac{|k_{i}^{\mathrm{tar}} - \bar{k}_{i}|}{k_{i}^{\mathrm{tar}}}.
\end{equation}
If $\max \left\lbrace \epsilon_{i} \right\rbrace_{i} < \eta$, where $\eta$ is a tolerance which we set to $\eta = 10^{-2}$, continue to step 3. Otherwise, go back to step 2c.
\end{itemize}

\item[3.] \textbf{Generating the network with the $\mathbb{S}^1$ model:} In this last step, we simply connect every pair of nodes with the probabilities in Eq.~\eqref{eq:connection_probability}. Since step 2 assigns a hidden degree to every node such that its expected degree matches its target degree, the resulting observed degrees in the network must be similar to the target degrees as well.
\end{itemize}
Figure~\ref{fig:angles} shows the networks generated by the model for different values of $\gamma$ and $\Lambda$. As in Ref.~\cite{Zuev:2015aa}, the angular distribution has an evident soft-community structure for low values of $\Lambda$, whereas for high values of the initial attractiveness, the angular density resembles that of the homogeneous $\mathbb{S}^1$ model. Despite the considerable differences in the similarity space distances between nodes for different values of $\Lambda$, the displayed networks are extremely similar from a topological perspective (see Fig.~\ref{fig:topologies}), with almost undistinguishable degree distributions and clustering spectra. Notice that step 2 is not specifically designed for the GPA angular distribution\footnote{The ordering of the target degrees might not be necessary in a more general situation where, for instance, hidden degrees are not correlated with angles.}. In principle, it should be valid for other distributions as well.

\section{Discussion}

There is abundant evidence of the geometric origin of many properties of complex networks, not only regarding their topology~\cite{Boguna2010,Serrano:2012we,gulyas2015navigable,Garcia-Perez:2016,Garcia-Perez:2017}, but also their weighted organisation~\cite{allard2017geometric}. The field of network geometry has therefore attracted much attention recently, and the $\mathbb{S}^1$ model is one of its cornerstones. On the one hand, it provides an intuitive and plausible explanation for clustering in real networks by introducing the concept of similarity space. On the other hand, it allows to build geometric maps of real networks by embedding them. These maps are remarkably meaningful, to the extent of predicting symmetries in real systems~\cite{Garcia-Perez:2017}. In addition, they are very useful; they can be used to navigate the network efficiently~\cite{Boguna2010}, to detect communities~\cite{Boguna2010,Garcia-Perez:2016} or even to construct smaller-scale replicas of real networks for efficiently testing dynamics on real networks~\cite{Garcia-Perez:2017}.

So far, the $\mathbb{S}^1$ model has only been studied under several simplifying premises, like power-law degree distributions or independent hidden variables. Yet, it has been able to explain many observed phenomena in complex networks. However, it can be exploited beyond these assumptions, since the correlation between hidden degrees and angles might clarify many more topological features of real-world networks. This work opens the path towards such line of study by showing that the model \textit{does not require} those simplifying assumptions, as it is capable of generating topologically similar networks with highly correlated hidden variables.

Moreover, the results presented in this paper might also have an impact on the embedding of real networks. Typically, the likelihood maximisation procedure only seeks the best angular coordinates, whereas hidden degrees are considered to be a function of degree only and known from the start~\cite{Boguna2010,Serrano:2012we}. This hypothesis is a direct consequence of the aforementioned simplifying assumptions usually contemplated in the $\mathbb{S}^1$ model. Nevertheless, as we have shown in this work, a heterogeneous angular distribution requires correcting hidden degrees in such a way that they depend on the hidden variables of all other nodes. This is a very important result, since it suggests that inferring hidden degrees via likelihood maximisation as well might noticeably improve the quality of embeddings of real-world networks with community structure.

\begin{acknowledgements}
We acknowledge support from a James S. McDonnell Foundation Scholar Award in Complex Systems; the ICREA Academia prize, funded by the Generalitat de Catalunya; Ministerio de Econom\'{\i}a y Competitividad of Spain projects no. FIS2013-47282-C2-1-P and no. FIS2016-76830-C2-2-P (AEI/FEDER, UE); the Generalitat de Catalunya grant no. 2014SGR608.
\end{acknowledgements}

\bibliographystyle{spphys}

\end{document}